# Organ size increases with obesity and correlates with cancer risk.

Haley Grant[1, ^], Yifan Zhang[1, ^], Lu Li[1, ^], Yan Wang[2], Satomi Kawamoto[3], Sophie Pénisson[4], Daniel F. Fouladi[3], Shahab Shayesteh[3], Alejandra Blanco[3], Saeed Ghandili[3], Eva Zinreich[3], Jefferson S. Graves[3], Seyoun Park[3], Scott Kern[5,6], Jody Hooper[6], Alan L. Yuille[2, 7], Elliot K Fishman[3], Linda Chu[3], Cristian Tomasetti[1, 5, 8, *]

^ These authors contributed equally.

* Corresponding author: ctomasetti@jhu.edu

---

[1] Department of Biostatistics, Johns Hopkins Bloomberg School of Public Health, 615 N Wolfe St, Baltimore, MD 21205, USA. Current address: Amazon, San Francisco, CA, USA.
[2] Department of Computer Science, Whiting School of Engineering, Johns Hopkins University, Baltimore, MD 21218, USA.
[3] Department of Radiology, Johns Hopkins School of Medicine, Baltimore, MD 21205, USA.
[4] Laboratoire d'Analyse et de Mathématiques Appliquées, Université Paris-Est Créteil, 94000 Créteil, France.
[5] Sidney Kimmel Comprehensive Cancer Center, Johns Hopkins School of Medicine, Baltimore, MD 21205, USA.
[6] Department of Pathology, Johns Hopkins School of Medicine, Baltimore, MD 21205, USA.
[7] Department of Cognitive Science, Krieger School of Arts and Sciences, Johns Hopkins University, Baltimore, MD 21218, USA.
[8] Division of Biostatistics and Bioinformatics, Department of Oncology, Johns Hopkins School of Medicine, Baltimore, MD 21205, USA.

# Abstract

Obesity increases significantly cancer risk in various organs. Although this has been recognized for decades, the mechanism through which this happens has never been explained. Here, we show that the volumes of kidneys, pancreas, and liver are strongly correlated (median correlation = 0.625; P-value<$10^{-47}$) with the body mass index (BMI) of an individual. We also find a significant relationship between the increase in organ volume and the increase in cancer risk (P-value<$10^{-12}$). These results provide a mechanism explaining why obese individuals have higher cancer risk in several organs: the larger the organ volume the more cells at risk of becoming cancerous. These findings are important for a better understanding of the effects obesity has on cancer risk and, more generally, for the development of better preventive strategies to limit the mortality caused by obesity.

Obesity, as defined by a body mass index (BMI) greater than 30, is now considered the most important lifestyle factor contributing to cancer (1). The incidence of obesity is increasing worldwide, and the number of cancers impacted by it in the world is huge (2-5). As a consequence, obesity is the focus of an enormous amount of research and funding today, arguably the most studied risk factor in cancer. Notwithstanding all the past and current efforts, the mechanisms underlying the effects of obesity on cancer risk have proved to be elusive.

For example, when comparing the mutational loads of obese versus non-obese patients with a given cancer type, using sequencing data available in the The Cancer Genome Atlas (TCGA) database, often no large differences are found (6). That is, obesity does not cause an increase in the overall mutation rate. And harmful inherited mutations appear to explain only a very small percentage (<5%) of obesity (7). This is somewhat surprising since cancer is by and large the result of deleterious genetic and epigenetic changes as described by the somatic mutation theory of cancer (8-10) and successively solidified by genome-wide analyses (11-16). How can obesity increase the risk of cancer if it does not increase a cell's mutational load?

There are certainly other well-recognized factors involved in carcinogenesis beyond an increase in mutation rate. For example, the microenvironment can have important influences on the growth of a tumor (17). A weakened immune system may facilitate a tumor to escape immune surveillance, even harnessing our body's natural defense to the tumor's advantage (18). Current research efforts to elucidate the mechanisms connecting obesity to cancer risk are therefore directed toward its inflammatory and metabolic effects, such as the effects of hormones like putative adipokines (19-21) or the effects of circulating metabolites on cell growth (22). However, no definitive cancer-promoting mechanism has yet been substantiated.

In our previous research, we showed the fundamental role played by the number of cells, and the number of cell divisions, in determining cancer risk (15, 23-25). A natural question is then to ask whether obesity has an effect on the size of our organs, and therefore on their total numbers of cells. The underlying hypothesis for a cancer-promoting mechanism is that obesity may induce an increase in the number of cells forming an organ. That increase would then lead to an increase of the risk of getting cancer in that organ, as cancer can originate from each of its cells at risk: a double number of those cells would yield a double risk. As it will be shown, our analyses indicate that this indeed appears to be the case in the three different organs we were able to analyze.

## Results

**Extrapolations of Organ Size at Varying BMI**

We started investigating the relationship between BMI and organ size by searching the literature for estimates of organ size wherein BMI annotation had been provided. As data were scarce and covered only relatively small BMI ranges, we used linear extrapolations to extend the analysis to a full range of BMI values. Encouraging results were obtained in all the seven different organs we were able to analyze: gallbladder, thyroid, liver, pancreas, kidney, gastric cardia, and corpus uteri (see Table S1 and Supplementary File). Often the extrapolations indicated an increase of 50% to a doubling of the organ size when going from normal weight (BMI = 20-25) to very extreme obesity (BMI=50).

**Measurements of Organ Size at Varying BMI via Imaging**

Given that these were only statistical extrapolations from limited data, we then decided to measure the variation in organ size among several hundreds of individuals and across a full range of BMI by analyzing a large set of imaging data. Specifically, the volume of kidneys, pancreas, and liver was measured in 750 individuals by expert radiologists via segmentation of computed tomography (CT) scan images. These subjects were selected from a much larger database of ~50,000 individuals so as to include only those without a known liver, pancreatic, or kidney disease, using exclusion criteria for patients having cancer or large (>5cm) cysts or other benign lesions, and requiring identical parameters with respect to the CT scan procedure used to produce the images. The solid organs of liver, kidneys, and pancreas were specifically chosen, as the volumes of these organs for a given subject will remain relatively constant during short intervals of time (i.e. hours or days), whereas the volumes of hollow organs such as stomach, small bowel, and colon will vary based on oral intake and degree of luminal distension.

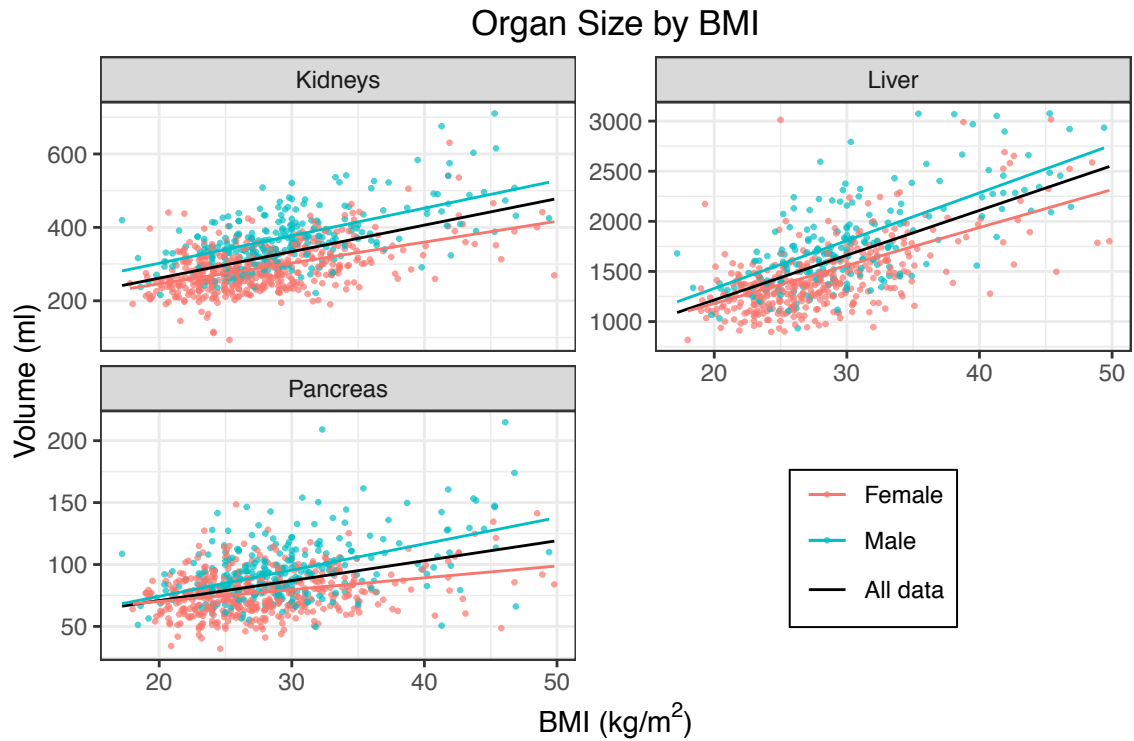

**Figure 1. Body mass index (BMI) versus organ volume.** Weighted linear regressions using BMI as a predictor for organ volume for kidneys (both left and right), liver, and pancreas. Lines correspond to the fitted regression line, one using all data points (black); one, only data from females (red); and one, only data from males (blue). Data points and lines are colored by gender.

Consistent with our initial extrapolations, we observed highly significant positive Pearson's correlations between BMI and organ size, not previously described (Figure 1, and Tables S2-S3). Specifically, the Pearson's correlations were 0.625 (95%CI: 0.57-0.68; P-value<$10^{-79}$), 0.74

(95%CI: 0.68-0.79; P-value<$10^{-101}$), and 0.51 (95%CI: 0.44-0.57; P-value<$10^{-47}$), in kidneys, liver, and pancreas, respectively.

We found that for every 5-point increase in BMI from baseline (BMI=18.5-24.9) the volumes of kidneys, liver, and pancreas increase by 11% (CI: 10%-12%), 13% (CI: 12%-14%), 8% (CI: 7%-9%), respectively. Patients with BMI close to 50 have organs that are generally between 50% and 100% larger in volume than patients with a healthy weight (see Figure 2 for an example).

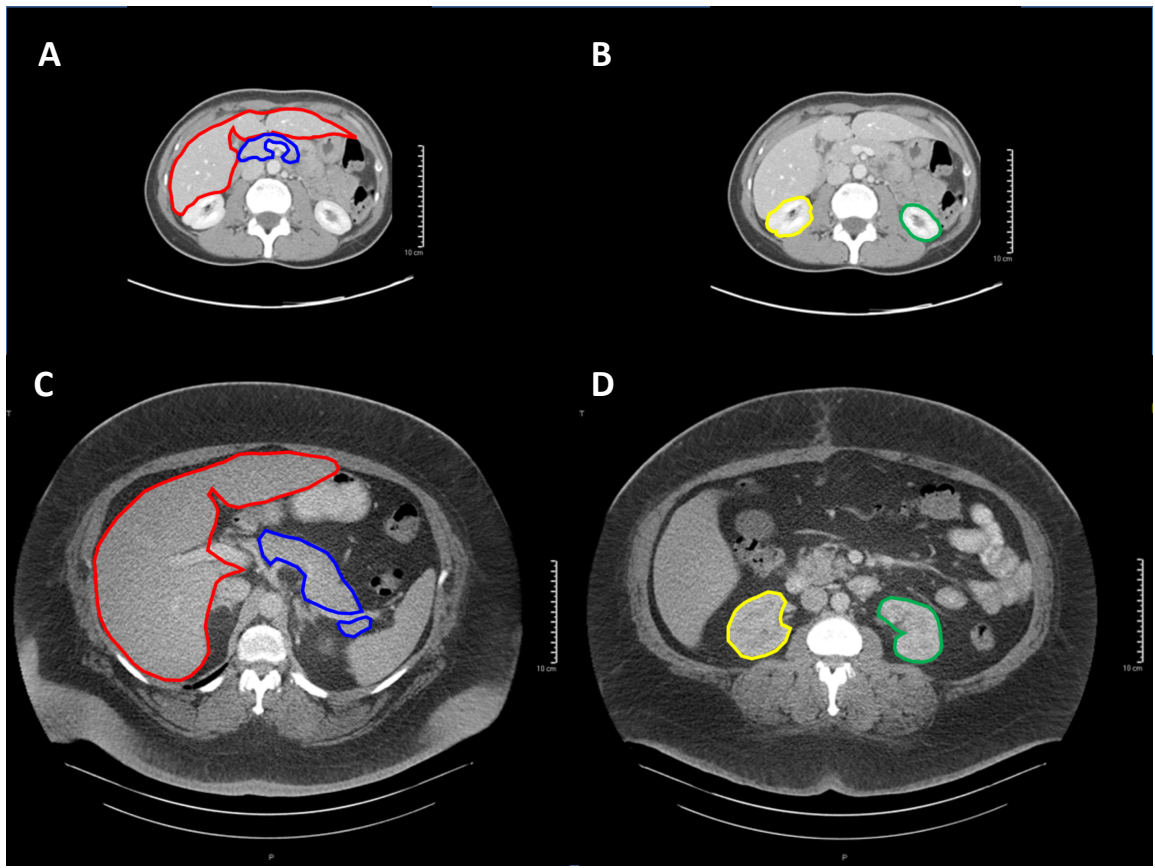

**Figure 2. Comparison of the organ volumes of a normal versus an obese patient.** Comparison of organs' volumes between a normal body weight individual (female, 45 years old, BMI=20.7) versus an extremely obese one (female, 48 years old, BMI=53.9). The contours of liver (red), pancreas (blue), and kidneys (yellow – left kidney; green – right kidney) in the normal-weight individual (A, B), versus the obese one (C, D) are depicted. Images have the same scale, as indicated.

To further validate our analyses, we considered a subset of CT scans, from 67 of the original 750 individuals, and reanalyze them by applying a recent automated image-segmentation technique that uses artificial intelligence (AI), employing deep learning with neural networks, to find the boundary of a given organ and its three-dimensional (3D) volume in CT scan images (see Supplementary File). All the manually segmented and computer predicted organ contours were verified by a board-certified abdominal radiologist. The results obtained with this orthogonal method confirmed our original findings, yielding comparable results (see Figure S1 and Table S3).

The increase in organ volume we have observed is neither affected by subcutaneous fat nor by visceral fat, given the imaging technique. Could it be an increase in the number of fat cells within the organ? Previous research indicated that only a relatively small proportion of this increase in organ volume may be due to either an increase in the volume of cells, as caused by a greater fat content, or an increase in the number of fat cells. Specifically it has been shown that, when comparing obese to normal weight individuals, the fat content stays approximately the same in kidneys (median 1.35% vs 0.64%), slightly increases in pancreas (3.60% vs 2.26%), and increases more substantially in liver (4.57% vs 1.11%) (26). Those increases, however, can only explain a very small fraction of the volume increases we have observed. And the number of adipocytes is rather tightly regulated during adulthood (for subjects with median BMI=40 the number of fat cells was less than double the amount found in the normal weight subjects) (27). If the number of fat cells or the volume of fat cells, or the fat content, can only explain a relatively small part of the large increase in organ volume of an obese person, it must be that these larger volumes must primarily reflect an expansion in the number of non-fat, parenchymal cells in those organs.

**Correlation Between the Increase in Organ Size and the Relative Risk of Cancer**

Given the well-known relationship between obesity and cancer risk (28), we then compared the observed increase in organ volume caused by obesity with the reported relative risk of cancer in that organ. We found a significant correlation for all three tissues: 0.89 (P-value<0.018), 0.74 (P-value<0.006), 0.80 (P-value<0.0017) for kidneys, liver, and pancreas, respectively (Table S4). For example, overweight people (BMI: 25-29.9) have 18% (95%CI: 11%-25%), 18% (95%CI: 12%-24%), and 12% (95%CI: 6%-19%) larger kidneys, liver, and pancreas, respectively, and a corresponding 28% (95%CI: 24%-33%), 17% (95%CI: 5%-30%), and 18% (95%CI: 3%-36%) higher cancer risk in those organs. Obese people (BMI ≥30) have 55% (95%CI: 46%-66%), 68% (95%CI: 59%-76%), and 39% (95%CI: 29%-49%) larger kidneys, liver, and pancreas, respectively, and a corresponding 77% (95%CI: 68% - 87%), 84% (95%CI: 56%-98%), and 47% (95%CI: 23%-75%) higher cancer risk (Table S4). When combining data of all three tissues, to assess the relationship between the relative increase in volume of patients in a given BMI interval against the corresponding relative risk of cancer, the coefficient of determination ($R^2$) was 0.77 (P-value<$10^{-12}$) (Figure 3). Therefore, almost 80% of the variation in cancer risk attributable to the effect of obesity can be explained by the variation that obesity induces in the parenchymal volume of the three organs. The linear regression depicted in Figure 3 has an estimated slope of 1.17 (P-value: <$10^{-12}$), implying that for an organ undergoing a doubling in volume, we expect to see an increase in cancer risk of 117% (i.e. about double). From a mathematical perspective this is not surprising, in fact this is what mathematical modeling of tumor evolution predicts, as the relationship between number of cells at risk and cancer risk is linear: a doubling in the number of cells at risk should yield a doubling of the lifetime risk of cancer in that organ (see Methods).

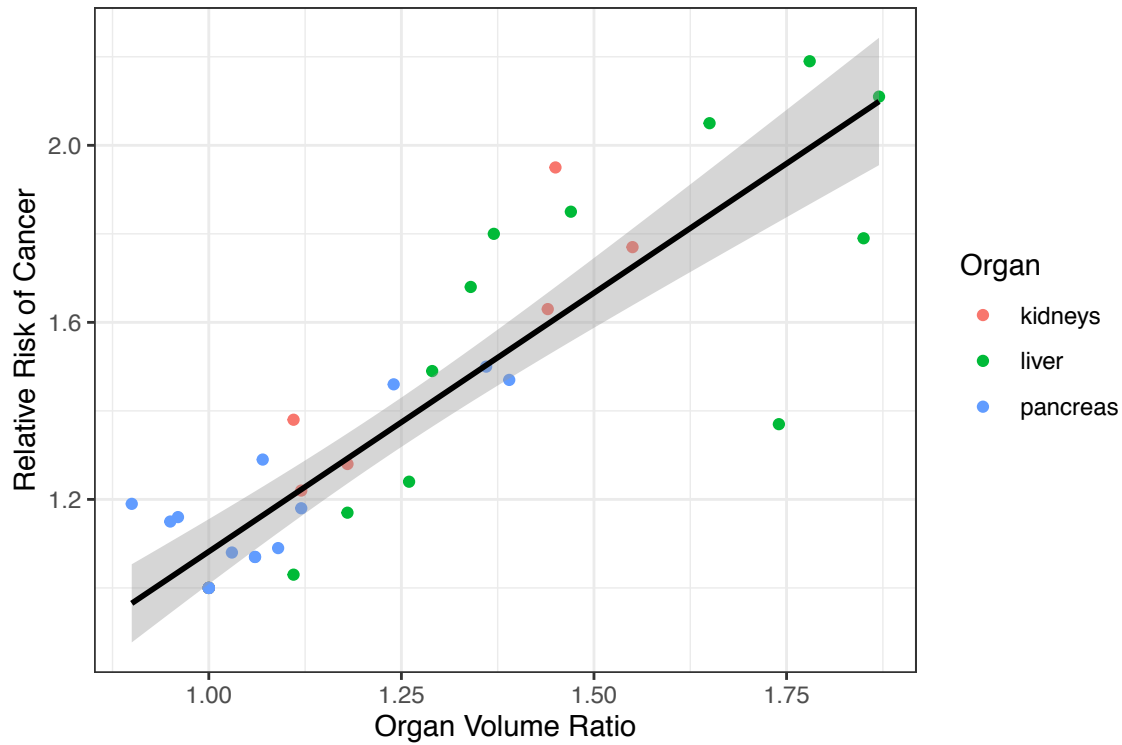

**Figure 3. Organ volume ratio versus cancer relative risk.** The relationship between the actual organ volume in a subject with respect to the volume at normal BMI (Organ Volume Ratio) and the corresponding relative risk of cancer, in kidneys (red), liver (green), and pancreas (blue), respectively. For each color, a point corresponds to a given gender and BMI interval. The estimated slope is equal to 1.17 (P-value: $<10^{-12}$).

## Discussion

We have experimentally shown that several organs in obese patients are significantly larger than normal, in a dose-response relationship with BMI. We have also shown a positive correlation and a linear relationship between an increase in organ size and its cancer relative risk. Overall, our findings provide a mechanism through which obesity increases cancer risk. They show that an increase in the number of cells at risk in an organ is associated with a higher risk of cancer in that organ. A major effect that obesity exerts on tumorigenesis is not an increase in the mutation rate, like smoking does to lung, but rather an increase in the number of cells that can carry those mutations.

These findings provide further support to the already available evidence that the number of cells in an organ play a critical role for its cancer risk, as previously shown across a large number of organs (24, 25). Large epidemiological studies over millions of individuals also support this mechanism, by showing that taller people have higher cancer risk (29, 30), with an estimated average 16-18% increase in overall cancer risk every 10 cm of extra height, and comparable estimates across each organ. This again indicates the important role that the number of cells plays in cancer risk since there are well-known, strong correlations between body height and organ size, e.g. a 1.5 cm increase in colon length - and the associated increase in number of cells - for every centimeter of extra body height (31, 32).

The relationship presented in Figure 3 provides more than just a correlation. It is striking that the estimated slope is approximately equal to one as this is precisely what is mathematically expected if the mechanism behind the increase in cancer risk is an increase in the number of cells at risk (Methods and (10)). If the slope of the relationship had been very different from one, the relative risk of cancer could not have been explained simply by the variation in organ size and the number of cells at risk.

Also, the relationship we have described is far from deterministic, as a large variation was found between BMI and organ size (Fig. 1), with some relatively normal BMI patients having larger-than-expected organs and vice versa. It will be important, then, to assess whether organ size may be a more accurate predictive marker of cancer risk than BMI, thus enhancing risk stratification and cancer prevention.

A limitation of our study is that we analyzed the effects of BMI only on three organs via imaging - and four more via mathematical extrapolations - for a total of seven out of the thirteen organs for which there is sufficient evidence of an effect of obesity on cancer risk, as evaluated by the International Agency for Research on Cancer (IARC) (see Table 2 in (28)). Thus, it will be important to extend our analyses to the remaining cancer types. Also, based on our extrapolations, the increase in volume does not appear sufficient to explain the extremely large increase of cancer risk observed in extremely obese women for the corpus uteri, indicating that other important factors may play key roles in that organ.

The measurements in our study were performed at one time point over the lifetime of a patient. An interesting question, which we leave to future research, is what effect weight loss - through diet or bariatric surgery - has on the size of various organs. Given some evidence that weight loss reduce cancer risk (33), we hypothesize that weight loss should reduce the size of those enlarged organs. And can an organ ever return to its normal size if BMI returns to its normal range? Answers to these questions will provide critical information for the primary prevention of cancer.

These results should have an impact on several research directions and preventive approaches. A better understanding of how obesity affects cancer risk and so many other diseases will help focus research and prevention efforts against being overweight and obesity, arguably among the greatest lifestyle problems in our contemporary society.

# Methods

**Extrapolations.** For Table S1, all methods, references, and approximations used are reported in the Supplementary Information.

**Segmentation.** The whole 3D volumes of liver, pancreas, and kidneys were manually segmented by six researchers using commercial segmentation software (Velocity 3.2.0, Varian Medical Systems) for each of the selected 750 individuals, who are part of a large database of patients who received a computed tomography (CT) scan at Johns Hopkins Medicine. The obtained contours were verified and approved by three abdominal radiologists with 5-30 years of experience. Patient's age, gender, and BMI at the time of the exam were obtained from the associated medical records.

**Data.** All volume measurement data obtained via manual segmentation are provided in Table S2. The table provides the patient ID (FELIX), age, gender, and volumes (arterial and venous) of liver, pancreas, and kidneys (right and left), respectively, for all 750 patients analyzed. For robustness, the value used in the analyses for the organ volume of each subject was the average between the arterial and the venous volumes.

**Statistical Analysis.** For Figure 1 all data are reported in Table S2. For Figure 1 and Table S3, weighted linear regressions were performed using BMI as a predictor for organ volume for kidneys (both left and right), liver, and pancreas. Weights were calculated by the inverse probability of observing a point in a 5 kg/m$^2$ intervals (BMI was split into 5 unit intervals within which all points received the same weight in the regression) to account for the imbalanced data structure as many more observations were in the 20-30 BMI range. Weighted correlations were taken using the same weights used in the weighted regression shown in Figure 1.

For Table S4, we found estimated relative risks (RR) of cancer in each organ type for different BMI intervals via a literature review: (34) for kidneys, (35) for liver, and (36) for pancreas. We then split our data into these same intervals and estimated the organ volume ratio in each organ compared to the organ volume in the normal BMI range. We used the interval [18.5, 25) as the normal range for liver and kidneys and [21,23) for pancreas because these were the intervals used for the meta-analyses in which cancer risks were reported. The organ volume ratio and corresponding 95% confidence interval were calculated using a bootstrap method in which we randomly sampled 100 observations from the interval in question and the baseline group (normal BMI), calculated the pairwise ratios within these samples, and then took the average ratio for each of 10000 iterations. The mean was our point estimate with a confidence interval taken to be the empirical .025 and .975 quantiles from the bootstrap. For the categories "30+" we imposed a further requirement that the sample be drawn equally across each 5 unit interval. That is, because of our data imbalance, with fewer observations falling in the high BMI ranges, we imposed balance in the bootstrap sampling by taking 25 observations from each of the 5-unit intervals (30-35, 35-40, 40-45, 45-50).

Once we obtained estimates of relative risk (RR) and the associated organ volume increase (Organ Volume Ratio) found in Table S4, we plotted each of those pairs of values - where each point corresponds to one BMI interval (for one gender category) – and combined all three organs in the same figure (Figure 3). Note that the reference categories (normal BMI) overlap at the point (1,1) on this plot by construction. A simple linear regression was fit using organ volume ratio as a predictor of relative risk across all three cancers and obtained an estimated coefficient equal to 1.17 (P-value: <10$^{-12}$).

**A.I. algorithm.** Alan Yuille's group has developed an automatic organ segmentation that uses training data to predict the organ contours in the validation dataset. The contours of the liver, kidneys, and pancreas were manually contoured on each CT slice in the training dataset. Automatic segmentation of the three internal organs (pancreas, liver, kidney) in the validation dataset was performed from CT images using organ attention networks, which first make a coarse segmentation for the organs, reducing the complex background, followed by a second stage analysis that refines target organs to a more accurate estimate. This work was published in MIA 2019. The performance of the algorithm is documented in (37) and is sufficiently accurate for segmenting organs, in terms of standard performance measures (e.g., DICE scores=98.0% for liver, 97.6% for kidney and 87.8% for pancreas).

**Mathematical relationship between number of cells at risk and cancer risk.** Probabilistically, a doubling in the number of cells at risk in an organ will yield a lifetime cancer risk equal to $1-(1-x)^2 = 2x-x^2$, where x is the original lifetime risk. However, the lifetime risk of cancer in a given organ is typically very small (often a few percentage points or smaller), which implies $x^2 \approx 0$, and therefore a doubling of the original risk (2x) is a good approximation.

**Acknowledgements**

CT would like to thank Bert Vogelstein for his mentorship and support. This work was supported by The John Templeton Foundation, The Lustgarten Foundation for Pancreatic Cancer Research, The Maryland Cigarette Restitution Fund, and NIH Grant P30-CA006973.

**Author Contributions**

CT conceived the idea and designed the overall project. ALY, EKF, LC, and CT designed the imaging analyses and provided supervision. HG, LL, and CT performed the statistical analyses. YZ and CT performed the mathematical extrapolations. SK, DFF, SS, AB, SG, EZ, JSG, SP, EKF, and LC performed the image segmentations. YW and ALY performed the A.I. analyses. SP, SK, and JD performed research. HG, YZ, LL, LC, and CT wrote the initial manuscript and all other authors contributed to the final version of the manuscript.

**Declaration of Interests**

Under a license agreement between Thrive and the Johns Hopkins University, C.T. and the University are entitled to royalty distributions. Additionally, the University owns equity in Thrive. C.T. is also a paid consultant to Amgen, Bayer AG, Johnson & Johnson, and Thrive. These arrangements have been reviewed and approved by the Johns Hopkins University in accordance with its conflict of interest policies.

| Cancer type | Gender | Relative organ size compared to BMI 22 | | | | | Relative risk of cancer* |
|---|---|---|---|---|---|---|---|
| | | BMI 30 | BMI 35 | BMI 40 | BMI 45 | BMI 50 | |
| Gallbladder | | 1.35 | 1.55 | 1.74 | 1.91 | 2.08 | 1.3 (1.2-1.4) |
| Thyroid | M | 1.28 | 1.46 | 1.63 | 1.81 | 1.98 | 1.1 (1.0 -1.1) |
| | F | 1.19 | 1.31 | 1,44 | 1.56 | 1.68 | |
| Liver | M | 1.38 | 1.62 | 1.86 | 2.09 | 2.33 | 1.8 (1.6-2.1) |
| | F | 1.36 | 1.58 | 1.8 | 2.02 | 2.24 | |
| Pancreas | | 1.09 | 1.15 | 1.21 | 1.27 | 1.33 | 1.5 (1.2-1.8) |
| Kidney | M | 1.26 | 1.42 | 1.59 | 1.75 | 1.91 | 1.8 (1.7-1.9) |
| | F | 1.28 | 1.46 | 1.63 | 1.81 | 1.98 | |
| Gastric cardia | | 1.41 | 1.63 | 1.84 | 2.04 | 2.23 | 1.8 (1.3-2.5) |
| Corpus uteri | F | 1.2 | 1.33 | 1.47 | 1.62 | 1.77 | 7.1 (6.3-8.1) |

**Table S1. Extrapolated relationship between BMI, organ size, and increase in cancer risk.** Organ volume estimated at different BMI values and compared with the corresponding relative risk (RR) of cancer. * RR for the BMI $\geq$ 40.0 category evaluated against normal BMI (18.5-24.9) (28). See Supplementary File 1 for the derivation of the relative organ size estimates.

**Table S2. Volume measurement data.** The table provides the patient ID (FELIX), age, gender, and volumes (arterial and venous) of liver, pancreas, and kidneys (right and left), respectively, for all 750 patients analyzed.

**Table S3. Correlation coefficients and corresponding P-values.** Estimated correlation coefficients and P-values for the analysis on kidneys (both left and right), liver, and pancreas, as depicted in Figure 1 (sheet 1). Regression coefficients and corresponding P-values for the validation analysis performed on kidneys (both left and right), liver, and pancreas on the original dataset (sheet 2) and the 67 additional patients, whose volumes were estimated from the images using an artificial intelligence (A.I.) algorithm (sheet 3).

**Table S4. Organ volume versus relative risk.** Observed increase in organ volume with respect to the normal BMI group (Organ Volume Ratio) compared to the corresponding relative risk (RR) of cancer in that organ for each BMI interval considered.

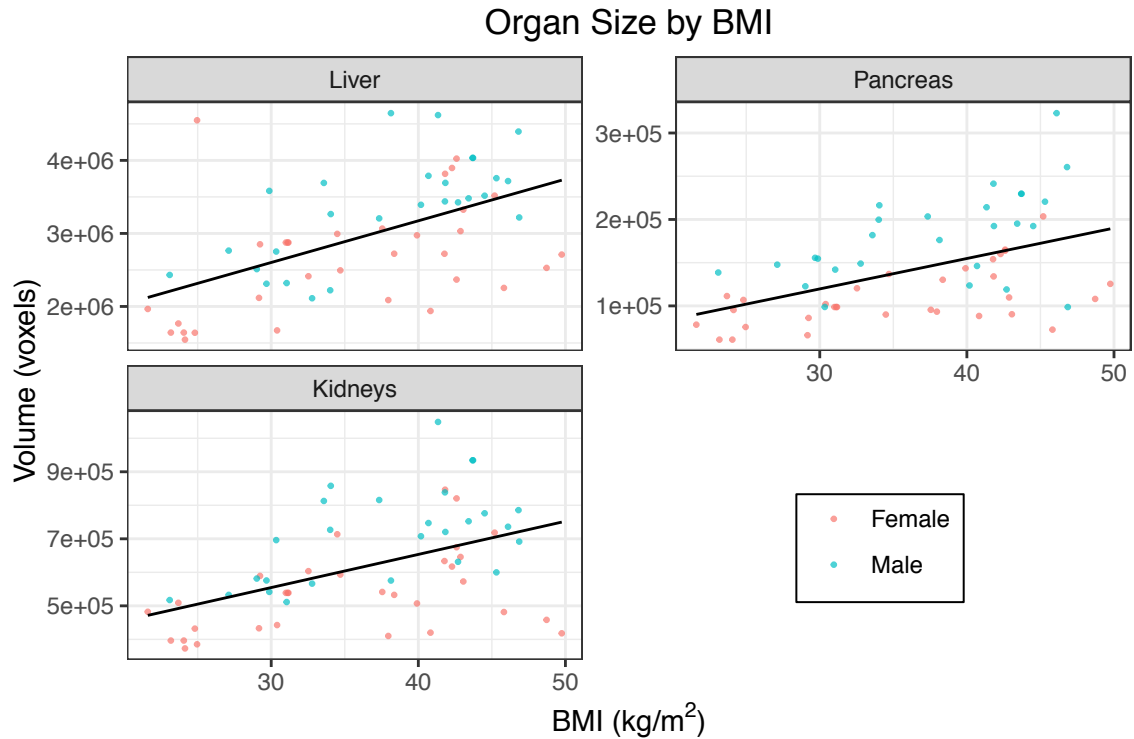

**Figure S1. Validation analysis on 67 cases.** Weighted linear regressions using BMI as a predictor for organ volume for kidneys (both left and right), liver, and pancreas for the validation set on 67 patients, whose volumes were estimated from the images using an artificial intelligence (A.I.) algorithm described in the Methods. Lines correspond to the fitted regression line, one using all data points (black), one using only data from females (red), and one only data from males (blue). Data points and lines are colored by gender.

# Supplementary Information

**Mathematical extrapolations of organ size as a function of BMI**

To explore how much of the increase in relative cancer risk can be explained by the increase in organ size alone, we wanted to compare the organ sizes of overweight and obese subjects with body mass index (BMI) = 30, 35, 40, 45, and 50 to organ sizes of subjects with a normal BMI of 22. Given the very limited availability of data on organ size at different BMI values, we linearly extrapolated from the limited data to obtain estimates across the full range of BMI and report those values in Table S1. In the following we explain how these extrapolations were obtained for each organ included in our analysis. We stress that, as explained in the main text, this was a preliminary analysis based on rough extrapolations whose objective was simply to determine if our hypothesis of a correlation between organ size and BMI was supported by the available data in the literature to justify the further, more detailed, investigations using CT scans that we then conducted and presented in the main text.

Gallbladder
In the study by Stone et al. (38) subjects with normal BMI ($22 \pm 1$) had an average gallbladder residual volume (after maximal emptying) of 4.2 ml compared to 8.4 ml for subjects with high BMI ($36 \pm 1$). Assuming a linear relationship between BMI and gallbladder residual volume, this translates into a 0.3 ml increase per BMI unit. A normal gallbladder measures around 7-10 cm in length and 2-3.5 cm in diameter with a wall thickness of 2-3 mm (39). We took the midpoints of these measurements and approximated the gallbladder as a spheroid, assuming that the relative proportions of the organ measurements stay the same with increases in organ size. Since nearly all gallbladder cancers begin in glandular cells that line the inner surface of the gallbladder, relative ratios of the organ inner surface area were calculated.

Thyroid
Sahin et al. provide regression equations relating thyroid volume to body weight separately for males and females (40). We are interested in BMI rather than weight but it is easy to link these two together as BMI is calculated as weight (in kilograms) divided by height (in meters) squared. Height was held constant at the average height in the U.S., which is 176 cm for men and 161.9 cm for women, while weights corresponding to different BMI values were recorded and used to estimate respective thyroid volumes. Ratios of thyroid volume of overweight and obese subjects compared to normal subjects were obtained.

Liver
In a study by Gallagher et al. (41), overweight and obese adults with type 2 diabetes were enrolled in a weight-loss intervention program. At the study baseline, the women weighed on average $82 \pm 14$ kg, and had an average liver weight of $1.76 \pm 0.50$ kg; the men weighed on average $93 \pm 8$ kg, and had an average liver weight of $1.90 \pm 0.40$ kg. After two years of weight-loss intervention, the subjects weighed on average 5.21 ± 0.74 kg less while MRI-derived measurements showed that their liver weighed 0.11 ± 0.03 kg less. This suggests a 0.02 kg decrease in liver weight per 1 kg decrease in body weight. Again, to get measurements in terms of BMI, corresponding weights at different BMI values were recorded with height held constant at the average American height. Since mass is proportional to volume, the ratios of liver masses corresponding to different BMI values are equal to the ratios of liver volumes corresponding to different BMI values.

Pancreas
Saisho et al. provide regression equations relating parenchymal pancreas volume to BMI (42). The information was used to obtain ratios of parenchymal pancreas volume of overweight and obese subjects compared to normal subjects.

Kidney
The most common type of kidney cancer is renal cell carcinoma, which originates in the lining of the proximal convoluted tubule. No information regarding effects of obesity on the dimensions of the proximal tubules could be found, so instead, here we assume that the surface area of the proximal tubules increases proportionally with kidney volume. In a study by Gallagher et al., overweight and obese adults with type 2 diabetes were enrolled in a weight-loss intervention program. At the study baseline, the women weighed on average 82 ± 14 kg, and had an average kidney weight of 0.38 ± 0.11 kg; the men weighed on average 93 ± 8 kg, and had an average kidney weight of 0.46 ± 0.08 kg (41). After two years of weight-loss intervention, the subjects weighed on average 5.21 ± 0.74 kg less while MRI-derived measurements showed that their kidney weighed 0.02 kg less. This suggests a 0.0038 kg decrease in kidney weight per 1 kg decrease in body weight. Again, to get measurements in terms of BMI, corresponding weights at different BMI values were recorded with height held constant at the average American height. The ratios of kidney masses corresponding to different BMI values are equal to the ratios of kidney volumes corresponding to different BMI values.

Gastric cardia
The stomach has five main regions, the cardia, the fundus, the body, the antrum, and the pylorus. No studies were found that specifically address the size of the gastric cardia in relation to BMI, so we assumed that the proportions of the regions of the stomach remain fixed. Or in other words, the size of the cardia increases proportionally with the size of the whole stomach. In the study by Geliebter, the mean gastric capacity for lean subjects (mean BMI = 22.3) was 1100 ml, and for obese subjects (mean BMI = 31.5) was 1925 ml (43). This translates into a 90 ml increase per BMI unit. Almost all stomach cancers are adenocarcinomas, which develop from cells that line the inside of the stomach. To compute surface area, the stomach is approximated as a spheroid. The average-sized human stomach has a volume of 0.94 L and a greater curvature that measures at 26-31 cm (44). Using 28.5 cm as the major axis length, the minor axis length is derived to be approximately 8 cm. These measurements allow us to define the shape of the stomach and thus estimate its surface area given its volume. Ratios of stomach surface area of overweight and obese subjects compared to normal subjects were obtained.

Corpus uteri
In the study by Parmar et al., the mean uterine length of parous women was 8.63 cm for those with a body weight 51-60 kg, and 9.06 cm for those with a body weight 61-70 kg (45). In nulliparous women, mean uterine length was 7.08 for those with a body weight 51-60 kg, and 7.45 cm for those with a body weight 61-70 kg. This translates into an increase in uterine length of 0.43 cm in parous women, and 0.37 cm in nulliparous women per 10 kg increase in body weight. According to the U.S. Census Bureau's Current Population Survey in 2018, 49.8 percent of women aged 15 to 44 had never had children (46). Taking this into account, the weighted increase in uterine length is 0.40 cm per 10 kg increase in weight. We approximate the corpus part of the uterus as a cone, where the ratio of the mean length to mean base diameter of the cavity of body of the uterus is taken to be 1.9 (47). Ratios of uterine cavity surface area of overweight and obese subjects compared to normal subjects were obtained.